\documentclass[twocolumn,english,showpacs,aps,pre,secnumarabic]{revtex4}
\usepackage[T1]{fontenc}
\usepackage[latin1]{inputenc}
\usepackage{amsmath}
\usepackage{graphicx}
\usepackage{amssymb}

\makeatletter

\providecommand{\tabularnewline}{\\}

\usepackage{amsfonts}

\usepackage{babel}

\usepackage{babel}
\makeatother
\begin{document}

\title{Master singular behavior from correlation length measurements for
seven one-component fluids near their gas-liquid critical point}

\author{Yves Garrabos{*}, Fabien Palencia, Carole Lecoutre, Can J. Erkey{*}{*}}

\affiliation{ESEME-CNRS, ICMCB-UPR 9048, Université Bordeaux 1, 87 avenue du Docteur
Albert Schweitzer, F-33608 Pessac France.}

\affiliation{{*}{*}on sabbatical leave from Department of Chemical Engineering,
University of Connecticut, Storrs, CT 06074 USA.}

\email{garrabos@icmcb-bordeaux.cnrs.fr}

\author{Bernard Le Neindre }

\affiliation{LIMHP-CNRS-UPR 1311, Université Paris 13, Avenue Jean Baptiste Clément,
F-93430 Villetaneuse France.}

\date{27 July 2005}

\begin{abstract}
We present the master (i.e. unique) behavior of the correlation length,
as a function of the thermal field along the critical isochore, asymptotically
close to the gas-liquid critical point of xenon, krypton, argon, helium
3, sulfur hexafluoride, carbon dioxide and heavy water. It is remarkable
that this unicity extends to the correction-to-scaling terms. The
critical parameter set which contains all the needed information to
reveal the master behavior, is composed of four thermodynamic coordinates
of the critical point and one adjustable parameter which accounts
for quantum effects in the helium 3 case. We use a scale dilatation
method applied to the relevant physical variables of the one-component
fluid subclass, in analogy with the basic hypothesis of the renormalization
theory. This master behavior for the correlation length satisfies
hyperscaling. We finally estimate the thermal field extent, where
the critical crossover of the singular thermodynamic and correlation
functions deviate from the theoretical crossover function obtained
from field theory. 
\end{abstract}

\pacs{64.60.-i, 05.70.Jk, 64.70.Fx}

\maketitle

\section{Introduction}

Close to the gas-liquid critical point of a one-component fluid, the
knowledge of the correlation length $\xi$, i.e. the size of the critical
fluctuations of the order parameter, is one among the most important
challenge to provide better critical phenomena understanding, in particular
for hyperscaling and crossover descriptions. The correlation length
measurements $\xi\left(\Delta T\right)$ as a function of the temperature
distance $\Delta T=T-T_{c}$ to the critical point along the critical
isochore $\rho=\rho_{c}$, in the homogeneous range $T>T_{c}$, have
been published for Xe \cite{Swinney 1973,Guttinger 1981}, Kr \cite{Bonetti 2003},
Ar \cite{Lin 1974}, $^{3}$He \cite{Ohbayashi 2003,Zhong 2003},
SF$_{6}$ \cite{Puglielli 1970,Cannell 1975}, CO$_{2}$ \cite{Maccabee 1971,Lunacek 1971},
and D$_{2}$O \cite{Sullivan 2000,Bonetti 2000}. $T$ ($T_{c}$)
is the temperature (critical temperature). $\rho$ ($\rho_{c}$) is
the density (critical density). We report an analysis of these data
using the scale dilatation method initially proposed by one of us
\cite{Garrabos 1982,Garrabos 1985}, which was recently upgraded \cite{Garrabos 2005a}
to account for quantum effects in light fluids such as $^{3}$He .
The basic information of the scale dilatation method is given by the
four coordinates which localize the critical point on the experimental
$p$ (pressure), $v_{\bar{p}}$ (particle volume), $T$ (temperature),
phase surface. In addition, a single well-defined adjustable parameter,
noted $\Lambda_{qe}^{*}$, characterizes the quantum contribution
at $T\cong T_{c}$. Considering such a minimum set of critical parameters,
the two main objectives of this paper are:

i) to phenomenologically observe the master (i.e. unique) singular
behavior of the correlation length for the universality subclass of
the one-component fluids. For this we simply use the appropriate scale
dilatation of $\Delta T$ and $\xi$, showing the master behavior
without exact knowledge of its singular functional form;

ii) to fit the resulting master curve by a mean crossover function
obtained from the recent results \cite{Bagnuls 2002} of the massive
renormalization scheme \cite{Bagnuls 1984a,Bagnuls 1985,Bagnuls 1987},
valid for the complete universality class of the three-dimensional
(3D) uniaxial symmetrical Ising like systems \cite{Zinn 2003}.

This paper is organized as follows. Section 2 provides the data sources.
Only are considered the effective fitting results of the correlation
length measurements which have been published in the litterature.
After the introduction of the four critical coordinates which characterize
each one-component fluid, we recall in Section 3 the essential features
of the singular behavior of the correlation length. From a brief analysis
based on the corresponding state scheme, we illustrate also the well-known
failure of the classical theories of critical phenomena \cite{Anisimov 2000}.
Section 4 presents the application of the scale dilatation method
to the physical (field) variables of the fluid subclass, leading to
the master singular behavior observed. A fitting (two-terms) power
law equation, which satisfies universal features of asymptotic hyperscaling
and (one-term) critical crossover valid in the preasymptotic domain
\cite{Bagnuls 1985}, is also proposed in this Section 4. Then the
fitting by a well-defined and complete mean crossover function obtained
from the massive renormalization scheme, is made in Section 5. A brief
analysis of the validity range of the classical-to-critical crossover
description is given before concluding in Section 6.

\section{The data sources}

The data sources are obtained from turbidity and scattering measurements
as a function of $T$, which provide simultaneous determination of
the susceptibility (proportional to the isothermal compressibility)
and the correlation length. The measurements are performed {}``\emph{near
the critical point}'', that corresponds to a finite temperature range
bounded by the max and min values of $\Delta T=T-T_{c,exp}$, where
$T_{c,exp}$ is the measured (or estimated) critical temperature in
the experiments. The relative precision estimated by the authors is
generally of the order of 10\%, but we have noted that the raw data
for $\xi$, as a function of the raw data for $\Delta T$, are scarcely
given in the published results to provide easy control of this uncertainty
level. The authors only have systematically reported their fitting
results as a function of the dimensionless temperature distance $\Delta\tau^{\ast}$
to the critical point, defined by\begin{equation}
\Delta\tau^{\ast}=\frac{\Delta T}{T_{c,exp}}=\frac{T-T_{c,exp}}{T_{c,exp}}\label{delta tau thermal field}\end{equation}
 Such a normalized temperature difference is the relevant physical
variable to describe the singular scaling behavior of the thermodynamic
fluid properties along the critical isochore \cite{Widom 1965}. In
the field theory framework \cite{Zinn 2003}, $\Delta\tau^{\ast}$
is proportional to the renormalized thermal field $t$ of the $\Phi_{d=3}^{4}\left(n=1\right)$-model
for the universality class considered in the present paper (see below)
which corresponds to a scalar ($n=1$) order parameter density and
a three dimensional ($d=3$) system.

The customary functional forms used to fit the data are:

i) the effective (single term) power law divergence \begin{equation}
\xi=\xi_{0}^{+}\left(\Delta\tau^{\ast}\right)^{-\nu}\label{ksi effective power law}\end{equation}
 where the free amplitude $\xi_{0}^{+}$ is a fluid-dependent quantity
and $\nu$ an effective critical exponent which only asymptotically
($\Delta\tau^{\ast}\rightarrow0$) takes a universal theoretical value,
estimated to $\nu_{Ising}\approx0.63$ (see \cite{Guida 1998} for
updated theoretical estimations). In Eq. (\ref{ksi effective power law}),
the value of the critical exponent can be considered as an adjustable
parameter when measurements are performed in a restricted temperature
range at finite distance to $T_{c,exp}$. The main characteristic
of such fitting is the high correlation between the effective values
of $\xi_{0}^{+}$ and $\nu$, which are then highly dependent on $T_{c,exp}$
and on the (extension and mean) values of the temperature range covered
by measurements.

ii) the Wegner expansion \cite{Wegner 1972}, generally restricted
to the following (two-term) equation,

\begin{equation}
\xi=\xi_{0}^{+}\left(\Delta\tau^{\ast}\right)^{-\nu}\left[1+a_{\xi}^{+}\left(\Delta\tau^{\ast}\right)^{\Delta}\right],\label{ksi wegner expansion}\end{equation}
 where $\Delta\approx0.5$ is a universal critical exponent \cite{Guida 1998}
which characterizes the leading family of the corrections to the scaling
behavior. $a_{\xi}^{+}$ is the fluid-dependent confluent amplitude
of the first correction to scaling. In such a fitting equation, the
exponents are generally fixed to their theoretical values, and only
the adjustable amplitudes $\xi_{0}^{+}$ and $a_{\xi}^{+}$ remain
highly correlated to $T_{c,exp}$. Moreover the contribution of the
first confluent correction term to scaling mostly appears lower than
- or of the same order of magnitude as - the experimental uncertainty
(reflecting an effective experimental situation where $a_{\xi}^{+}\left(>0\right)\sim1$
and $\Delta\tau^{\ast}\lesssim10^{-2}$).

In Table \ref{tab1} are summarized the selected fitting results {[}with
free (or fixed) exponents and accounting (or not) for first-order
Wegner term{]} for the seven fluids. All these fitting results are
taken from litterature (see references given in the last column of
Table \ref{tab1}). 

\begin{table*}
\begin{tabular}{|c|c|c|c|c|c|c|c|}
\hline 
\multicolumn{1}{|c|}{Fluid}&
 $\begin{array}{c}
\xi_{0}^{+}\\
\left(\textrm{Å}\right)\end{array}$&
 $\nu$&
 $a_{\xi}^{+}$&
 $\Delta$&
 $\begin{array}{c}
\Delta T_{max}\\
\left(\textrm{\textrm{K}}\right)\end{array}$&
 $\begin{array}{c}
\Delta T_{min}\\
\left(\textrm{\textrm{K}}\right)\end{array}$&
 Ref.\tabularnewline
\hline
$Xe$&
 $\begin{array}{c}
2.0\pm0.25\end{array}$&
 $\begin{array}{c}
0.63\pm0.05\end{array}$&
&
&
&
&
 \cite{Swinney 1973}\tabularnewline
\hline
$Xe$&
 $\begin{array}{c}
1.84\pm0.03\end{array}$&
 $\begin{array}{c}
0.63\\
fixed\end{array}$&
 $0.55$&
 $\begin{array}{c}
1/2\\
fixed\end{array}$&
 $\begin{array}{c}
10\end{array}$&
$0.0026$&
 \cite{Guttinger 1981}\tabularnewline
\hline
$Kr$&
 $\begin{array}{c}
1.71\pm0.01\end{array}$&
 $\begin{array}{c}
0.6304\\
fixed\end{array}$&
 $\begin{array}{c}
0.624\pm0.4\end{array}$&
 $\begin{array}{c}
0.504\\
fixed\end{array}$&
 $\begin{array}{c}
20\end{array}$&
$0.021$&
 \cite{Bonetti 2003}\tabularnewline
\hline
$Ar$&
 $\begin{array}{c}
1.71\pm0.25\end{array}$&
 $\begin{array}{c}
0.63\pm0.02\end{array}$&
&
&
 $\begin{array}{c}
3.5\end{array}$&
$0.04$&
 \cite{Lin 1974}\tabularnewline
\hline
$Ar$&
 $\begin{array}{c}
1.6\pm0.2\end{array}$&
 $\begin{array}{c}
0.64\pm0.02\end{array}$&
&
&
 $\begin{array}{c}
3.5\end{array}$&
$0.04$&
 \cite{Lin 1974}\tabularnewline
\hline
$^{3}He$&
 $\begin{array}{c}
4.8\pm2.0\end{array}$&
 $\begin{array}{c}
0.59\pm0.04\end{array}$&
&
&
 $\begin{array}{c}
0.00014\end{array}$&
$0.000014$&
 \cite{Ohbayashi 2003}\tabularnewline
\hline
$^{3}He$&
 $\begin{array}{c}
2.71\pm0.015\end{array}$&
 $\begin{array}{c}
0.629\pm0.002\end{array}$&
 $\begin{array}{c}
0.732\pm0.007\end{array}$&
 $\begin{array}{c}
0.502\pm0.01\end{array}$&
 $\begin{array}{c}
0.1\end{array}$&
$0.000014$&
 \cite{Zhong 2003}\tabularnewline
\hline
$SF_{6}$&
 $\begin{array}{c}
1.5\pm0.23\end{array}$&
 $\begin{array}{c}
0.67\pm0.07\end{array}$&
&
&
 $\begin{array}{c}
0.45\end{array}$&
$0.038$&
 \cite{Puglielli 1970}\tabularnewline
\hline
$SF_{6}$&
 $\begin{array}{c}
2.016\pm0.2\end{array}$&
 $\begin{array}{c}
0.6214\pm0.01\end{array}$&
&
&
 $\begin{array}{c}
1.0\end{array}$&
$0.048$&
 \cite{Cannell 1975}\tabularnewline
\hline
$CO_{2}$&
 $\begin{array}{c}
1.94\pm0.2\end{array}$&
 $\begin{array}{c}
0.60\pm0.02\end{array}$&
&
&
 $\begin{array}{c}
10\end{array}$&
$0.01$&
 \cite{Maccabee 1971}\tabularnewline
\hline
$CO_{2}$&
 $\begin{array}{c}
1.50\pm0.09\end{array}$&
 $\begin{array}{c}
0.633\pm0.01\end{array}$&
&
&
 $\begin{array}{c}
10\end{array}$&
$0.023$&
 \cite{Lunacek 1971}\tabularnewline
\hline
$D_{2}O$&
 $\begin{array}{c}
1.30\pm0.23\end{array}$&
 $\begin{array}{c}
0.623\pm0.03\end{array}$&
&
&
 $\begin{array}{c}
13\end{array}$&
$0.15$&
 \cite{Sullivan 2000}\tabularnewline
\hline
$D_{2}O$&
 $\begin{array}{c}
1.372\pm0.01\end{array}$&
 $\begin{array}{c}
0.6304\\
fixed\end{array}$&
 $\begin{array}{c}
0.676\pm0.2\end{array}$&
 $\begin{array}{c}
0.504\\
fixed\end{array}$&
 $\begin{array}{c}
22.\end{array}$&
$1.6$&
 \cite{Bonetti 2000} \tabularnewline
\hline
\end{tabular}

\caption{Published values of $\xi_{0}^{+}$, $\nu$, $a_{\xi}^{+}$, and $\Delta$,
obtained from fitting {[}with the Eqs. (\ref{ksi effective power law},\ref{ksi wegner expansion}){]}
the turbidity and scattering measurements, in the temperature range
$\Delta T_{\min}\leqslant T-T_{c}\leqslant\Delta T_{\max}$, along
the critical isochore of seven one-component fluids (for data sources
and the selected fitting results see the references given in the last
column). \label{tab1}}
\end{table*}

\section{Analysis from the corresponding state scheme}

In Figure 1a (log-log scale; color online), are reported the curves
illustrating the fitted singular behavior of the $\xi\,\left(nm\right)$
raw data, as a function of $T-T_{c}\,\left(K\right)$ (from Eqs. (\ref{ksi effective power law})
or (\ref{ksi wegner expansion}) and data of Table 1). Each curve
has an extension covering the experimental temperature range (while,
at the Figure 1a scale, the curve thickness illustrates the 10\% uncertainty
on $\xi$ measurements). The dimensional quantities make each fluid
behavior clearly distinguishable (at the same value of $T-T_{c}=40\, mK$
for example, the $\xi$ values are covering one decade: from $5\, nm$
for $^{3}$He, up to $50\, nm$ for D$_{2}$O).

\begin{figure*}
\includegraphics[%
  width=150mm,
  keepaspectratio]{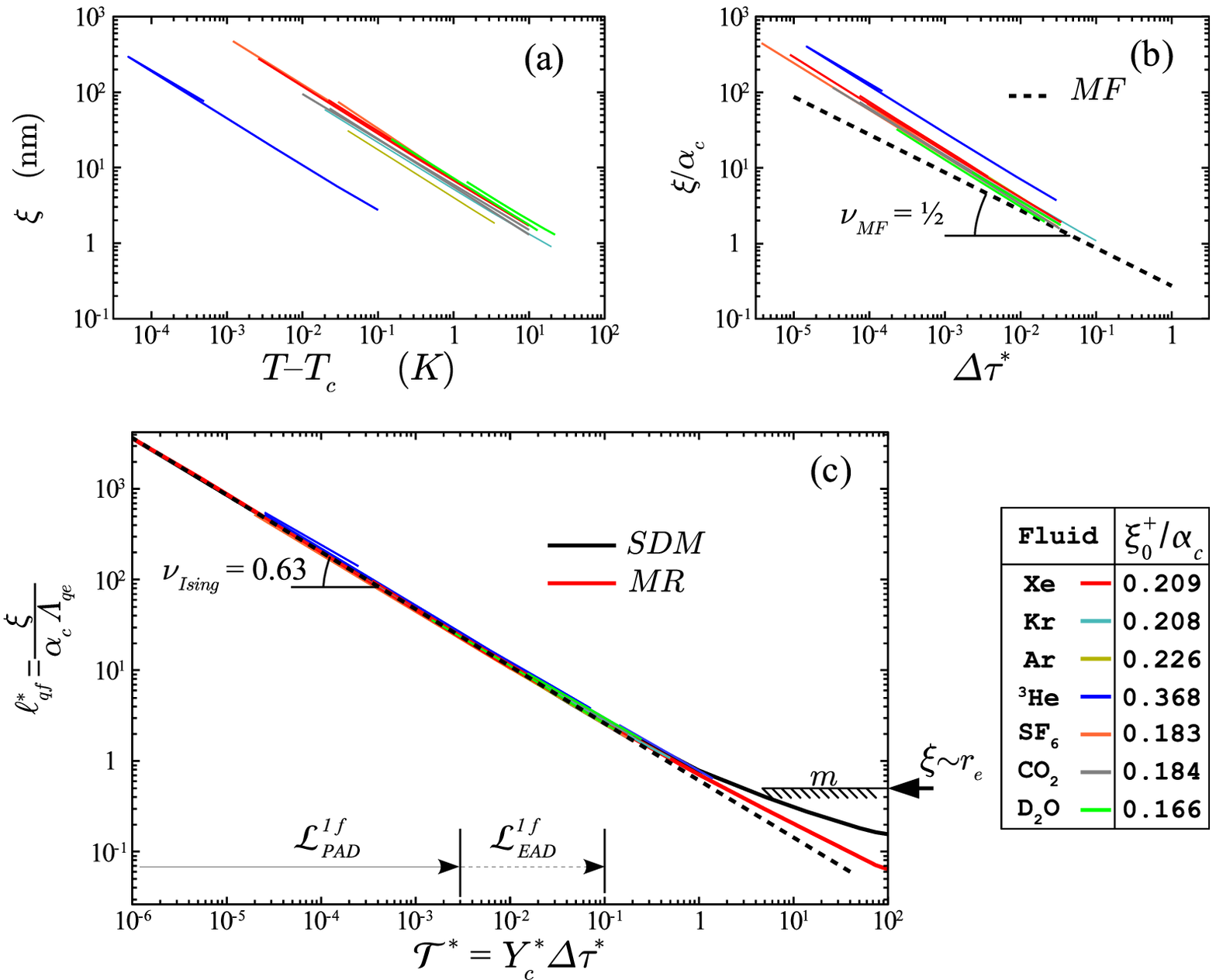}

\caption{(Color online) a) Log-Log scale of $\xi$ as a function of $T-T_{c}>0$,
along the critical isochore for Xe, Kr, Ar, $^{3}$He, SF$_{6}$,
CO$_{2}$, and D$_{2}$O. Each corresponding curve corresponds to
the fit results reported in Table I, using Eqs. (\ref{ksi effective power law})
and (\ref{ksi wegner expansion}) for fitting data on the corresponding
experimental range; b) Log-Log scale of the respective dimensionless
variables expressed in $\alpha_{c}$ length unit and $\left(\beta_{c}\right)^{-1}$
energy unit, illustrating the failure of the classical corresponding
state scheme. The expected mean field behavior with \emph{classical}
index $\nu_{MF}=\frac{1}{2}$ is given by the curve labelled $MF$
{[}see text and Eq. (\ref{ksi mean field}){]}; c) Matched master
behavior (Log-Log scale) of the \emph{renormalized} correlation length
$\ell_{qf}^{*}$, as a function of the \emph{renormalized} thermal
field $\mathcal{T}^{*}$, {[}see Eqs. (\ref{ksi dilatation}) and
(\ref{delta tau dilatation}), respectively{]}. The curves labelled
$SDM$ and $MR$ correspond to the Eqs. (\ref{two-term master equation})
and (\ref{ksi fitting equation}), respectively. The dashed line illustrates
the universal \emph{critical} index $\nu_{Ising}\simeq0.63$ for the
asymptotic pure power law behavior of the uniaxial 3D Ising-like universality
class. At large values of the renormalized thermal field ($\mathcal{T}^{*}\geq3$),
we have illustrated by the curve $m$ a rough estimate $\ell_{qf}^{*}\approx\frac{1}{2}$
of the \emph{microscopic} limit where the correlation length $\xi$
will reach the order of magnitude of the two particle equilibrium
position $r_{e}$ (with $r_{e}\gtrapprox\sigma$, where $\sigma$
is the \emph{size} of the particle); see text. In the superimposed
Table are shown color indexation for each fluid (left column) and
calculated $\frac{\xi_{0}^{+}}{\alpha_{c}}$ values (right column),
using Eq. (\ref{ksizero versus MR amplitude}) and Table II data. }
\end{figure*}

Our first analysis starts from the following characteristic set \cite{Garrabos 1982,Garrabos 1985},
\begin{equation}
Q_{c,a_{\bar{p}}}^{min}=\left\{ T_{c},v_{\bar{p},c},p_{c},\gamma_{c}^{\prime}=\left[\left(\frac{\partial p}{\partial T}\right)_{v_{\bar{p},c}}\right]_{CP}\right\} \label{qcmin critical coordinates}\end{equation}
 made of the four critical parameters needed to localize the critical
point on the $p$, $v_{\bar{p}}$, $T$ phase surface. The selected
data are given in Table II. The subscript $c$ corresponds to the
critical parameters, the subscript $a_{\bar{p}}$ recall the thermodynamic
potential (see below) used to construct its associated phase surface,
while the subscript $\bar{p}$ corresponds to a quantity normalized
per particle. The subscript $CP$ means the value at the critical
point. $v_{\bar{p}}=\frac{V}{N}=\frac{m_{\bar{p}}}{\rho}$, where
$V$ is the total volume of the fluid container, and $N$ is the total
amount of fluid particles of individual mass $m_{\bar{p}}$ \cite{particle mass}.
Such a phase surface of equation $\Phi_{a_{\bar{p}}}^{p}\left(p,v_{\bar{p}},T\right)=0$
uses the two conjugated experimental variables, $p$ (intensive) and
$V$ (extensive), and represents the fluid equilibrium states provided
by the equation of state (e.o.s.) $p\left(T,v_{\bar{p}}\right)=-\left(\frac{\partial A}{\partial V}\right)_{T,N}=-\left(\frac{\partial a_{\bar{p}}}{\partial v_{\bar{p}}}\right)_{T}$
\cite{Modell 1983}. $A\left(\Omega_{i}\right)$ {[}$a_{\bar{p}}\left(\omega_{i}\right)=\frac{A}{N}${]}
is the total (per particle) Helmholtz free energy of natural variables
$\Omega_{i}=\left(T,V,N\right)$ {[}$\omega_{i}=\left(T,v_{\bar{p}}\right)${]}.
$\gamma_{c}^{\prime}$ is the common limiting direction at CP, in
the $p;T$ plane, of both the critical isochore on the homogeneous
(single phase) domain ($T>T_{c}$), and the saturation pressure curve
- {[}the projection of the vapor-liquid equilibrium state{]} - on
the heterogeneous (two phase) domain ($T<T_{c}$). We note that the
$T_{c}$ values given in Table II, which result from complete thermodynamic
analysis of the phase surface, are mostly different from $T_{c,exp}$,
as the $\rho_{c}=\frac{m_{\bar{p}}}{v_{\bar{p},c}}$ values from Table
II generally differ from the measured (or estimated) critical density
in the selected experiments.

\begin{table*}
\begin{tabular}{|c|c|c|c|c|c|c|c|c|c|c|}
\cline{2-2} \cline{3-3} \cline{4-4} \cline{5-5} \cline{6-6} \cline{7-7} \cline{8-8} \cline{9-9} \cline{10-10} \cline{11-11} 
\multicolumn{1}{|c|}{Fluid}&
 $\begin{array}{c}
m_{\bar{p}}\\
\left(10^{-26}kg\right)\end{array}$&
 $\Lambda_{qe}^{\ast}$&
 $\begin{array}{c}
T_{c}\\
\left(K\right)\end{array}$&
 $\begin{array}{c}
v_{\bar{p},c}\\
\left(nm^{3}\right)\end{array}$&
 $\begin{array}{c}
p_{c}\\
\left(MPa\right)\end{array}$&
 $\begin{array}{c}
\gamma_{c}^{\prime}\\
\left(MPa\, K^{-1}\right)\end{array}$&
 $\begin{array}{c}
\left(\beta_{c}\right)^{-1}\\
\left(10^{-21}J\right)\end{array}$&
 $\begin{array}{c}
\alpha_{c}\\
\left(nm\right)\end{array}$&
 $Z_{c}$&
 $Y_{c}$\tabularnewline
\hline
$Xe$&
 $21.803$&
 $1$&
 $289.74$&
 $0.19589$&
 $5.84$&
 $0.118$&
 $4.0003$&
 $0.881508$&
 $0.28601$&
 $4.85434$\tabularnewline
\hline
$Kr$&
 $13.9154$&
 $1$&
 $209.286$&
 $0.15292$&
 $5.5$&
 $0.1562$&
 $2.88951$&
 $0.806901$&
 $0.291065$&
 $4.94372$\tabularnewline
\hline
$Ar$&
 $6.6336$&
 $1$&
 $150.725$&
 $0.12388$&
 $4.865$&
 $0.172$&
 $2.08099$&
 $0.753463$&
 $0.2896$&
 $4.32882$\tabularnewline
\hline
$^{3}He$&
 $0.4983$&
 $1.11966$&
 $3.31555$&
 $0.12022$&
 $0.114724$&
 $0.11759$&
 $0.0457761$&
 $0.736198$&
 $0.301284$&
 $2.39837$\tabularnewline
\hline
$SF_{6}$&
 $24.252$&
 $1$&
 $318.70$&
 $0.32684$&
 $3.76$&
 $0.0835$&
 $4.4$&
 $1.054$&
 $0.281$&
 $6.08$\tabularnewline
\hline
$CO_{2}$&
 $7.308$&
 $1$&
 $304.14$&
 $0.15622$&
 $7.3753$&
 $0.170$&
 $4.2$&
 $0.829$&
 $0.274$&
 $6.01$\tabularnewline
\hline
$D_{2}O$&
 $3.329$&
 $1$&
 $643.89$&
 $0.09346$&
 $21.671$&
 $0.2717$&
 $8.88987$&
 $0.743029$&
 $0.227829$&
 $7.07277$ \tabularnewline
\hline
\end{tabular}

\caption{Minimal set of critical parameters {[}see Eqs. (\ref{qcmin critical coordinates})
and (\ref{qcmin scale factors}){]} for the seven one-component fluids
of particle mass $m_{\bar{p}}$. $\Lambda_{qe}^{\ast}\geq1$ {[}Eq.
(\ref{Lambdaqestar}){]} differs from unity by a nonuniversal adjustable
quantity proper to the nature of the $^{3}He$ quantum particle {[}see
text and Eqs (\ref{qf wave number}) and (\ref{lambdac}){]} \label{tab2}}
\end{table*}

In a first step, from $Q_{c,a_{\bar{p}}}^{\min}$, we are able to
derive dimensionless thermodynamic and correlation functions, using
the scale factors, \begin{equation}
\left(\beta_{c}\right)^{-1}=k_{B}T_{c}\label{energy scale}\end{equation}
 as an energy unit, and\begin{equation}
\alpha_{c}=\left(\frac{k_{B}T_{c}}{p_{c}}\right)^{\frac{1}{d}}\label{length scale}\end{equation}
 as a length unit. In regards to the noticeable differences in the
$\xi$ values observed in Figure 1a, at a same $T-T_{c}$ value, we
note the small differences in $\alpha_{c}$ values, given in the column
8 of Table II. We also note that $\alpha_{c}$, obtained only using
intensive variables, is not dependent of the size $L\sim\left(V\right)^{\frac{1}{d}}$
of the container ($k_{B}$ is the Boltzmann constant; $d=3)$. $\alpha_{c}$
has a clear physical meaning as a length unit \cite{Garrabos 1982}:
it represents the spatial extent of the short-ranged (Lennard-Jones
like) molecular interaction \cite{Hirschfelder 1964}, which allows
us to define $v_{c,I}=\frac{k_{B}T_{c}}{p_{c}}$ as the volume of
the \textit{\emph{microscopic}} \textit{critical interaction cell}
of each fluid.

The introduction of the two characteristic dimensionless numbers,\begin{equation}
Z_{c}=\frac{p_{c}v_{\bar{p},c}}{k_{B}T_{c}}\label{critical compression factor}\end{equation}
 and\begin{equation}
Y_{c}=\gamma_{c}^{\prime}\frac{T_{c}}{p_{c}}-1\label{critical isochoric factor}\end{equation}
 allows us to rewrite the minimal set {[}Eq. (\ref{qcmin critical coordinates}){]}
in the more convenient form\begin{equation}
Q_{c,a_{\bar{p}}}^{\min}=\left\{ \left(\beta_{c}\right)^{-1},\alpha_{c},Z_{c},Y_{c}\right\} \label{qcmin scale factors}\end{equation}
 which involves one energy scale factor, one length scale factor,
and two dimensionless \emph{scale factors} characterizing two preferred
directions to cross the critical point along the critical isotherm
and the critical isochore, respectively. $Z_{c}$ is the usual critical
compression factor. In addition, $\left(Z_{c}\right)^{-1}=n_{c}v_{c,I}$
is the number of particles that fill $v_{c,I}$, and \emph{the minimal
set given by the equation} (\ref{qcmin scale factors}) \emph{appears
related to the critical interaction cell properties}.

We recall that the critical compression factor $Z_{c}$, and the critical
Riedel factor $\alpha_{R,c}$ - related to $Y_{c}$ by $\alpha_{R,c}=Y_{c}+1$
-, are also two basic tools for developing an e.o.s. for engineering
fluid modeling.

Figure 1b (log-log scale; color online), gives the singular behavior
of the dimensionless correlation length $\xi^{*}=\frac{\xi}{\alpha_{c}}$
as a function of the dimensionless temperature distance $\Delta\tau^{\ast}$
of Eq. (\ref{delta tau thermal field}), precisely obtained from the
classical theory of corresponding states (here with two characteristic
parameters), using $\left(\beta_{c}\right)^{-1}$ and $\alpha_{c}$
units. Evaluation from standard mono-atomic Xe, shows the failure
of the classical theory, \textit{increasing} quantum effects in $^{3}$He,
and \textit{decreasing} non-spherical interaction effects in D$_{2}$O.
The dimensionless correlation length is then covering a relative variation
by a factor more than 2 at the same reduced temperature distance to
the critical point (see also our calculated values of $\frac{\xi_{0}^{+}}{\alpha_{c}}$
in the inserted table on Figure 1). Moreover an illustration of the
inaptness of the {}``mean field'' exponent $\nu_{MF}=\frac{1}{2}$
to describe the \textit{classical} fluctuation behavior \cite{Kostrowicka 2004}
expected from Van der Waals-like theories, is shown by the curve labelled
$MF$, of equation\begin{equation}
\frac{\xi_{MF}}{\alpha_{c}}=\frac{1}{4}\left(\frac{3\sqrt{2}}{\pi}\right)^{\frac{1}{d}}\left(\Delta\tau^{\ast}\right)^{-\nu_{MF}}\label{ksi mean field}\end{equation}
 We note also that the classical approach along the critical isochore
in the homogeneous domain is not able to reproduce experimental behavior
at large distance to the critical point. Such a failure can be related
to the number of interacting particles (here given by $\frac{1}{Z_{c}}\sim3-4$),
which seems too small to validate a {}``mean field approximation''
of the attractive molecular interaction at distance greater than $\alpha_{c}$
for a fluid at critical density, as we will discuss below (see Section
5).

\section{Analysis using the scale dilatation method}

As demonstrated in references \cite{Garrabos 1985,Garrabos 2002},
$Z_{c}$ and $Y_{c}$ can be used as the two scale factors to formulate
the dimensionless master asymptotic behavior of the one component
fluid subclass. As a matter of fact, asymptotic master singular behaviors
of dimensionless potentials and dimensionless correlation functions
only occur \cite{Garrabos 1982} using appropriate dilatations of
the following physical variables, \begin{equation}
\Delta\tau^{\ast}=k_{B}\beta_{c}\left(T-T_{c}\right)\label{thermal field}\end{equation}
 (to generate the \emph{renormalized} thermal field),\begin{equation}
\Delta h^{\ast}=\beta_{c}\left(\mu_{\bar{p}}-\mu_{\bar{p},c}\right)\label{ordering field}\end{equation}
 (to generate the \emph{renormalized} ordering field), and\begin{equation}
\Delta m^{\ast}=\left(\alpha_{c}\right)^{d}\left(n-n_{c}\right)\label{OP density}\end{equation}
 (to generate density of the \emph{renormalized} order parameter -
conjugated to the \emph{renormalized} ordering field). These scale
dilatations of the fluid variables are now defined \cite{Garrabos 2005a}
in complete formal analogy to the field theory framework \cite{Zinn 2003}.
Such a theoretical approach provides a comprehensive understanding
of the diverging universal character of the spontaneous fluctuations
of extensive variables, using the renormalization group (RG) techniques
\cite{Wilson 1971,Wilson 1974,Bagnuls 2000}, to deal with the contributions
of \textit{critical} fluctuations. We briefly recall that this theoretical
approach accounts for infinite degrees of freedom near the non-Gausssian
(Wilson-Fisher) fixed point, throughout the Landau-Ginzburg-Wilson
Hamiltonian of the $\Phi_{d=3}^{4}\left(n=1\right)$-model for the
universality class of the 3D uniaxial symmetrical Ising like systems,
with associated coupling constant $u_{4}>0$. In this model, the relevant
pair of renormalized fields are the weakly fluctuating thermal field
$t$, and the ordering field $h$ which exhibits stronger fluctuations,
with \{$t=0$,$h=0$\} at the isolated non-Gaussian fixed point. A
single parameter $\kappa$, such that $\kappa<<\Lambda_{0}$, measures
the distance to the non-Gaussian fixed point in such a way that this
fixed point corresponds to $\kappa^{\ast}=\frac{\kappa}{\Lambda_{0}}=0$.
$\Lambda_{0}$ is the actual microscopic wave number, characterizing
a discrete structure of matter with spacing $\left(\Lambda_{0}\right)^{-1}$.
$\kappa$ is precisely related to the inverse correlation length $\xi^{-1}$
of the fluctuations of the order-parameter $m$, conjugated to $h$,
with $\ell_{qf}^{\ast}=\Lambda_{0}\xi=\left(\kappa^{\ast}\right)^{-1}$.
Here $\ell_{qf}^{\ast}$ corresponds to the dimensionless form of
the actual fluid correlation length expressed in units of $\alpha_{c}$
{[}Eq. (\ref{length scale}){]}, also including quantum fluids (labelled
$qf$) throughout the introduction of a dimensionless adjustable parameter
$\Lambda_{qe}^{\ast}$ which account for the quantum effects {[}see
below the Eq. (\ref{ksi dilatation}) and the related discussion{]}.
At the fixed point, the fluctuations are \textit{infinite}, i.e. $\xi\left(t=0,h=0\right)\sim\infty$.
Close to the non-Gaussian fixed point within the critical asymptotic
domain, i. e. for small values of the renormalized fields $t$ and
$h$ which ensure that $\kappa<<\Lambda_{0}$, the Wegner expansions
\cite{Wegner 1972} represent the singular behavior of thermodynamics
and correlations functions of any physical system belonging to the
universality class of this $\Phi_{d=3}^{4}\left(n=1\right)$-model.
In particular, the Wegner expansion of Equation (\ref{ksi wegner expansion})
can be used for $\xi$. In that critical asymptotic domain, $\xi$,
although \textit{finite}, is still larger than $\left(\Lambda_{0}\right)^{-1}$.
So that the close critical vicinity of the non-Gaussian fixed point
can be defined by $\ell_{qf}^{\ast}=\Lambda_{0}\xi>>1$.

The renormalization introduces the two-scale universality of the physical
system through analytical proportionality between the physical variables
and the renormalized fields $t$ and $h$, respectively \cite{Wilson 1971,Wilson 1974}.
In a similar manner, the scale dilatation method is defined by the
following \emph{renormalization} of $\Delta\tau^{*}$ and $\Delta h^{\ast}$
into $\mathcal{T}_{qf}^{\ast}$ and $\mathcal{H}_{qf}^{\ast}$ 

\begin{equation}
\mathcal{T}_{qf}^{\ast}\equiv\mathcal{T}^{\ast}=Y_{c}\Delta\tau^{*}\label{delta tau dilatation}\end{equation}

\begin{equation}
\mathcal{H}_{qf}^{\ast}=\left(\Lambda_{qe}^{\ast}\right)^{2}\mathcal{H}^{\ast}=\left(\Lambda_{qe}^{\ast}\right)^{2}\left(Z_{c}\right)^{-\frac{d}{2}}\Delta h^{\ast}\label{delta h dilatation}\end{equation}
 respectively. Correspondingly, the \emph{renormalization} of the
order parameter $\Delta m^{\ast}$ into $\mathcal{M}_{qf}^{\ast}$
reads as follows

\begin{equation}
\mathcal{M}_{qf}^{\ast}=\Lambda_{qe}^{\ast}\mathcal{M}^{\ast}=\Lambda_{qe}^{\ast}\left(Z_{c}\right)^{\frac{d}{2}}\Delta m^{\ast}\label{delta m dilatation}\end{equation}
 Obviously, $\mathcal{T}^{\ast}$, $\mathcal{H}^{\ast}$ and $\mathcal{M}^{\ast}$,
are the renormalized variables defined for non-quantum fluids, for
which the nonuniversal wave number reads $\Lambda_{0}=\frac{1}{\alpha_{c}}$,
since $\alpha_{c}$ is the single explicit length unit. In Eq. (\ref{delta tau dilatation}),
the identity $\mathcal{T}_{qf}^{\ast}\equiv\mathcal{T}^{\ast}$ means
that the quantum effects are only accounted for at $T\cong T_{c}$.
In Eqs. (\ref{delta h dilatation}) and (\ref{delta m dilatation}),
the dimensionless parameter\begin{equation}
\Lambda_{qe}^{\ast}=1+\lambda_{c}\label{Lambdaqestar}\end{equation}
 accounts for quantum effects on the microscopic wave number $\Lambda_{0}$
at $T\cong T_{c}$ \cite{Garrabos 2005a}, in such a relative phenomenological
way that\begin{equation}
\Lambda_{0}\Lambda_{qe}^{\ast}=\frac{1}{\alpha_{c}}\label{qf wave number}\end{equation}
 with\begin{equation}
\lambda_{c}=\lambda_{q,f}\frac{\Lambda_{T,c}}{\alpha_{c}}\label{lambdac}\end{equation}
 $\lambda_{q,f}$ (with $\lambda_{q,f}>0$), is thus a nonuniversal
adjustable number which accounts for statistical contribution due
to the nature (boson, fermion, etc.) of the quantum particle. $\Lambda_{T,c}=\frac{h_{P}}{\left(2\pi m_{\bar{p}}k_{B}T_{c}\right)^{\frac{1}{2}}}$
is the de Broglie thermal wavelength at $T=T_{c}$, $h_{P}$ is the
Planck constant (the subscript $P$ is here added to make a distinction
with the field theory ordering field $h$). The Eq. (\ref{qf wave number})
preserves the same length scale unit for thermodynamic and correlations
functions. 

Therefore, the \emph{renormalized} dimensionless correlation length
is given by\begin{equation}
\ell_{qf}^{\ast}=\left(\kappa^{\ast}\right)^{-1}=\Lambda_{0}\xi=\frac{\xi^{\ast}}{\Lambda_{qe}^{\ast}}=\frac{\xi}{\alpha_{c}\Lambda_{qe}^{\ast}}\label{ksi dilatation}\end{equation}
 and the corresponding dilatation of the dimensionless axis $\Delta\tau^{*}$
and $\frac{\xi}{\alpha_{c}}$ into $\mathcal{T}^{\ast}$ and $\ell_{qf}^{\ast}$
are then defined by Eqs. (\ref{delta tau dilatation}) and (\ref{ksi dilatation}),
respectively, only using $Q_{c,a_{\bar{p}}}^{\min}$ and $\Lambda_{qe}^{\ast}$. 

The expected collapsing onto the master behavior obtained from application
of this scale dilatation method to the physical variables is shown
in Figure 1c (log-log scale; color online), \emph{independently of
any theoretical form used to represent this master behavior}. The
scatter between the curves now corresponds to their estimated precision
(10\%) for each fluid correlation length.

Since the scale dilatation of the physical variables {[}see equations
(\ref{delta tau dilatation}) and (\ref{delta h dilatation}){]} is
analogous to the basic hypotheses of the renormalization group, we
expect that the master asymptotic singularities present the universal
features of the universality class. Specially within the preasymptotic
domain, the observed divergence of $\ell_{qf}^{\ast}$ can be represented
by the following two-term Wegner expansion

\begin{equation}
\ell_{qf}^{\ast}=\mathcal{Z}_{\xi}^{+}\left(\mathcal{T}^{\ast}\right)^{-\nu}\left[1+\mathcal{Z}_{\xi}^{1,+}\left(\mathcal{T}^{\ast}\right)^{\Delta}\right]\label{two-term master equation}\end{equation}
 where the leading amplitude $\mathcal{Z}_{\xi}^{+}=0.57$ and the
first confluent amplitude $\mathcal{Z}_{\xi}^{1,+}=0.377$ have \emph{master}
(i.e. constant) values for the pure fluid subclass (see the Refs.
\cite{Garrabos 1985,Garrabos 2002} for details making reference to
critical xenon behavior \cite{Bagnuls 1984b} in order to obtain these
master values). This result, obtained from master singular behavior
of thermodynamic properties satisfies asymptotic hyperscaling and
extends the scaling assertions first proposed by Widom \cite{Widom 1965}
for the equation of state of the one-component fluid.

\section{Mean crossover function from the massive renormalization scheme}

We can now consider the {}``min'' and {}``max'' accurate expressions
of the complete classical-to-critical crossover recently proposed
by Bagnuls and Bervillier \cite{Bagnuls 2002}. From the numeric values
of the parameters of the generic functions $F_{min}\left(t\right)$
and $F_{max}\left(t\right)$ given in Tables I and II of reference
\cite{Bagnuls 2002}, we have derived the numerical values of parameters
associated to the mean crossover functions which reproduce as closely
as possible the error treatment initially made by the authors from
their generic functions. Such a mean crossover function for the inverse
correlation length reads as follows\begin{equation}
\left[\ell^{\ast}\left(t\right)\right]^{-1}=\mathbb{Z}_{\xi}^{+}\left(t\right)^{\nu}{\displaystyle \prod_{i=1}^{i=3}}\left[1+X_{\xi,i}t^{D\left(t\right)}\right]^{Y_{\xi,i}}\label{mean crossover function}\end{equation}
 with\begin{equation}
D\left(t\right)=\frac{\Delta+\Delta_{MF}S_{2}\sqrt{t}}{1+S_{2}\sqrt{t}}\label{mean crosover exponent}\end{equation}
 and\begin{equation}
t=\vartheta\left|\Delta\tau^{*}\right|\label{renormalized thermal field}\end{equation}
 All the critical exponents ($\nu$, $\Delta$, $\Delta_{MF}$) and
the constants ($\mathbb{Z}_{\xi}^{+}$, $X_{\xi,i}$, $Y_{\xi,i}$,
$S_{2}$) are given in Table III. The adjustable parameter $\vartheta$
introduces the non-universality proper to each selected system.

\begin{table}
\begin{tabular}{|c|c|c|c|c|c|c|}
\hline 
\multicolumn{1}{|c|}{}&
 exponent&
 $\mathbb{Z}_{\xi}^{+}$&
 $S_{2}$&
 $i$&
 $X_{\xi,i}$&
 $Y_{\xi,i}$\tabularnewline
\hline
$\nu$&
 $0.6303875$&
 $2.121008$&
 $22.9007$&
 $1$&
 $40.0606$&
 $-0.098968$\tabularnewline
\hline
$\Delta$&
 $0.50189$&
&
&
 $2$&
 $11.9321$&
 $-0.15391$\tabularnewline
\hline
$\Delta_{MF}$&
 $0.5$&
&
&
 $3$&
 $1.90235$&
 $-0.00789505$\tabularnewline
\hline
&
&
&
&
&
 $\mathbb{Z}_{\xi}^{1,+}$&
 $5.81623$ \tabularnewline
\hline
\end{tabular}

\caption{Values of the universal exponents and constant parameters of Eqs.
(\ref{mean crossover function}), (\ref{mean crosover exponent})
and (\ref{MR confluent amplitude}). \label{tab3}}
\end{table}

To fit each curve of Figure 1a with equations (\ref{mean crossover function})
to (\ref{renormalized thermal field}), we introduce one asymptotic
(system-dependent) prefactor $\lambda_{0}$, of dimension $\left[length\right]^{-1}$,
from the following equation \begin{equation}
\frac{1}{\xi\left(\Delta\tau^{*}\right)}=\lambda_{0}\mathbb{Z}_{\xi}^{+}\left(\Delta\tau^{*}\right)^{\nu}{\displaystyle \prod_{i=1}^{i=3}}\left[1+X_{\xi,i}t^{D\left(t\right)}\right]^{Y_{\xi,i}}\label{ksi fitting equation}\end{equation}
 where we note that the leading term is now a unique function of $\Delta\tau^{*}$,
as proposed by Bagnuls and Bervillier \cite{Bagnuls 2002}. In such
a fitting procedure, $\vartheta$ is readily seen as a crossover parameter
associated with one irrelevant physical field. The prefactor $\lambda_{0}$
satisfies to the two-scale universal feature of this universality
class, associated with the two relevant physical fields (only two
among all these prefactors are characteristics of the non-universality
of the selected system). From our definition of the length unit, we
can rewrite $\lambda_{0}$ as follows\begin{equation}
\lambda_{0}=\frac{1}{\alpha_{c}}\ell_{0}^{*}\label{dimensionless prefactor}\end{equation}
 leading to the following modification of the above equation (\ref{ksi fitting equation})\begin{equation}
\frac{\alpha_{c}}{\xi\left(\Delta\tau^{*}\right)}=\ell_{0}^{*}\mathbb{Z}_{\xi}^{+}\left(\Delta\tau^{*}\right)^{\nu}{\displaystyle \prod_{i=1}^{i=3}}\left[1+X_{\xi,i}t^{D\left(t^{\ast}\right)}\right]^{Y_{\xi,i}}\label{ksistar fitting equation}\end{equation}

Now it is easy to understand that the restricted analysis of the two-term
Wegner expansion provides complete materials for the unequivocal determination
of the two adjustable parameters $\ell_{0}^{*}$ and $\vartheta$.
For example, the asymptotic term to term comparison of equation (\ref{ksi wegner expansion})
and inverse equation (\ref{ksistar fitting equation}), provides the
following two relations

\begin{equation}
\xi_{0}^{+}=\alpha_{c}\left(\ell_{0}^{*}\mathbb{Z}_{\xi}^{+}\right)^{-1}\label{ksizero versus MR amplitude}\end{equation}
 and

\begin{equation}
a_{\xi}^{+}=\mathbb{Z}_{\xi}^{1,+}\left(\vartheta\right)^{\Delta}\label{akis versus MR confluent amplitude}\end{equation}
 where \begin{equation}
\mathbb{Z}_{\xi}^{1,+}=-\sum_{i=1}^{i=3}X_{\xi,i}Y_{\xi,i}\label{MR confluent amplitude}\end{equation}
 The value of the constant amplitude $\mathbb{Z}_{\xi}^{1,+}$ is
given in Table III. However, we note that the quantum correction parameter
disappears in such a \emph{standard} estimation of the leading prefactor
(and leading amplitude) when the two scale factors are unknown \cite{Bagnuls 1984b}.

>From the scale dilatation method, to fit the master curve of Figure
1c with equations (\ref{mean crossover function}) to (\ref{renormalized thermal field}),
need to introduce two master (i.e. constant) parameters, $\ell_{0}^{\left\{ 1f\right\} }$
and $\vartheta^{\left\{ 1f\right\} }$, that are characteristics of
the one-component fluid subclass, using the crossover modeling equation
\begin{equation}
\begin{array}{rcl}
\frac{1}{\ell_{qf}^{\ast}\left(\mathcal{T}^{*}\right)} & = & \ell_{0}^{\left\{ 1f\right\} }\mathbb{Z}_{\xi}^{+}\left(\vartheta^{\left\{ 1f\right\} }\mathcal{T}^{*}\right)^{\nu}\\
 &  & \times{\displaystyle \prod_{i=1}^{i=3}}\left[1+X_{\xi,i}\left(\vartheta^{\left\{ 1f\right\} }\mathcal{T}^{*}\right)^{D\left(\vartheta^{\left\{ 1f\right\} }\mathcal{T}^{*}\right)}\right]^{Y_{\xi,i}}\end{array}\label{lqf fitting equation}\end{equation}
 In equation (\ref{lqf fitting equation}),

\begin{equation}
\ell_{0}^{\left\{ 1f\right\} }=\left[\mathcal{Z}_{\xi}^{+}\mathbb{Z}_{\xi}^{+}\left(\vartheta^{\left\{ 1f\right\} }\right)^{\nu}\right]^{-1}\label{l1fzero for  MR-master link}\end{equation}
 and

\begin{equation}
\vartheta^{\left\{ 1f\right\} }=\left(\frac{\mathbb{\mathcal{Z}}_{\xi}^{1,+}}{\mathbb{Z}_{\xi}^{1,+}}\right)^{\frac{1}{\Delta}}\label{teta1f for MR-master link}\end{equation}
 in order to agree with the two-term asymptotic behavior given by
equation (\ref{two-term master equation}). The mandatory relation
between the relevant field $t$ of the $\Phi_{d=3}^{4}\left(1\right)$-model
and the master field $\mathcal{T}^{*}$ of the fluid subclass reads
as follows\begin{equation}
t=\vartheta^{\left\{ 1f\right\} }\mathcal{T}^{*}\label{MR-master link}\end{equation}
 From $\mathbb{Z}_{\xi}^{+}$, $\mathbb{Z}_{\xi}^{1,+}$, $\mathcal{Z}_{\xi}^{+}$,
and $\mathbb{\mathcal{Z}}_{\xi}^{1,+}$ values we obtain $\vartheta^{\left\{ 1f\right\} }=0.004288$
and $\ell_{0}^{\left\{ 1f\right\} }=25.699$. We note that the \emph{master
prefactor} $\ell_{0}^{\left\{ 1f\right\} }$ is attached to the correlation
length behaviors above and below the critical temperature, while the
\emph{master crossover parameter} $\vartheta^{\left\{ 1f\right\} }$
is the same for any property along the critical isochore, above and
below the critical temperature. The respective curves labelled $SDM$
{[}of equation (\ref{two-term master equation}){]} and $MR$ {[}of
inverse equation (\ref{lqf fitting equation}){]}, are illustrated
in Figure 1c, with noticeable asymptotic agreement with master experimental
behavior of the seven one-component fluids. The preasymptotic domain
(labelled $PAD$), described by a Wegner expansion restricted to the
first confluent correction {[}see equation (\ref{two-term master equation}){]},
extends up to $\mathcal{L}_{PAD}^{1f}\lesssim3\,10^{-3}$ (see the
corresponding arrow in $\mathcal{T}^{*}$ axis). In the extended asymptotic
domain (labelled $EAD$) which extends up to $\mathcal{L}_{EAD}^{1f}\lesssim0.1$
(see the corresponding arrow in $\mathcal{T}^{*}$ axis), the observed
master behavior is well-represented by the theoretical critical-to-classical
crossover {[}see equation (\ref{lqf fitting equation}){]}. From the
comparison of Figures 1b and 1c, we can also note that the applicability
of the two-scale master behavior obtained from the scale dilatation
method goes far beyond the applicability of the corresponding state
method based on classical theory.

By reversing the scale dilatation method for any one-component fluid
where $Q_{c,a_{\bar{p}}}^{\min}$ and $\Lambda_{qe}^{\ast}$ are known,
it is easy to determine its attached two characteristic parameters
$\ell_{0}^{*}$ and $\vartheta$, and to derive Eq. (\ref{ksi wegner expansion})
from Eq. (\ref{two-term master equation}), using the following relations

\begin{equation}
\ell_{0}^{*}=\frac{1}{\Lambda_{qe}^{\ast}}\ell_{0}^{\left\{ 1f\right\} }\label{lzero for  MR-master link}\end{equation}

\begin{equation}
\vartheta=Y_{c}\vartheta^{\left\{ 1f\right\} }\label{teta for MR-master link}\end{equation}

\begin{equation}
\xi_{0}^{+}=\alpha_{c}\Lambda_{qe}^{\ast}\left(Y_{c}\right)^{-\nu}\mathcal{Z}_{\xi}^{+}\label{ksizero versus master amplitude}\end{equation}

\begin{equation}
a_{\xi}^{+}=\mathcal{Z}_{\xi}^{1,+}\left(Y_{c}\right)^{\Delta}\label{aksi versus master confluent amplitude}\end{equation}
Equation (\ref{ksi wegner expansion}) can now be readily used without
any adjustable parameter (except $\Lambda_{qe}^{\ast}$).

Since the critical behaviour predicted by RG theory asymptotically
agrees with the master behavior for the one-component fluid subclass,
the only remaining problem is the determination of the thermal field
distance at which significant deviation between the $\Phi_{d=3}^{4}\left(n=1\right)$-model
and the fluid subclass appears. To point out that such a thermal distance
(noted $\mathcal{T}_{CO}^{*}$ in the following) exists, we consider
the variation of the effective exponent, $\nu_{eff}=-\frac{dln\left(\ell_{qf}^{\ast}\right)}{dln\mathcal{T}}$,
as a function of $\mathcal{T}^{*}$. In addition, it is also possible
to consider the variation of the effective exponent, $\gamma_{eff}=-\frac{dln\left(\mathcal{X}_{qf}^{\ast}\right)}{dln\mathcal{T}}$
, which is now entirely known \cite{Garrabos 2005b}, thanks to universal
features predicted by RG theory and definitions {[}see equations (\ref{delta tau dilatation}),
(\ref{delta h dilatation}), and (\ref{delta m dilatation}){]} of
the scale dilatation. $\mathcal{X}_{qf}^{\ast}\left(\mathcal{T}\right)=\left(\frac{\partial\mathcal{M}_{qf}^{\ast}}{\partial\mathcal{H}_{qf}^{\ast}}\right)_{\mathcal{T}}=\left(\Lambda_{qe}^{\ast}\right)^{2-d}\left(Z_{c}\right)^{d-2}\kappa_{T}^{\ast}$
is the renormalized susceptibility \cite{Garrabos 2005a}, with $\kappa_{T}^{\ast}=p_{c}\kappa_{T}$.
$\kappa_{T}=\frac{1}{\rho}\left(\frac{\partial\rho}{\partial T}\right)_{T}$
is the (physical) isothermal compressibility. The two limits for the
$\gamma_{eff}$-variation calculated from the $\Phi_{d=3}^{4}\left(1\right)$-model
are $\gamma_{MF}=1$ (close to the Gaussian fixed point) and $\gamma_{Ising}\cong1.24$
\cite{Guida 1998} (close to the Wilson-Fisher fixed point).

The complete results are shown in Figure 2 (color online) as a function
of the renormalized thermal field $\mathcal{T}^{*}$ (lower axis),
or as a function of the renormalized correlation length $\ell_{qf}^{\ast}$
(upper axis). We recall that $\ell_{qf}^{\ast}$ gives a best estimate
of the ratio between the effective size ($\xi$) of the critical fluctuations
and the effective size ($\alpha_{c}$) of the attractive molecular
interaction. Since the typical range of the dispersion forces in Lennard-Jones-like
fluids is slighly greater than twice the equilibrium distance ($r_{e}\gtrapprox\sigma$)
between two interacting particles of finite core size $\sigma$, we
have reported on the upper axis of Figure 2 a rough estimate {[}$\ell_{qf}^{\ast}\approx\frac{1}{2}${]}
of the limit where the correlation length will be comparable with
the order of magnitude of the particle size. From the theoretical
crossover fit, such a limit corresponds to $\mathcal{T}_{\bar{p}}^{*}\approx1.3$
(here the supscript $\bar{p}$ is for a related particle property).
Therefore, it is important to note that the effective crossover for
the fluid subclass appears in the thermal field range $\mathcal{T}_{CO}^{*}\approx0.5-1$
where $\ell_{qf}^{\ast}\lesssim1$, as discussed below. 

\begin{figure}
\includegraphics[%
  width=1\columnwidth,
  keepaspectratio]{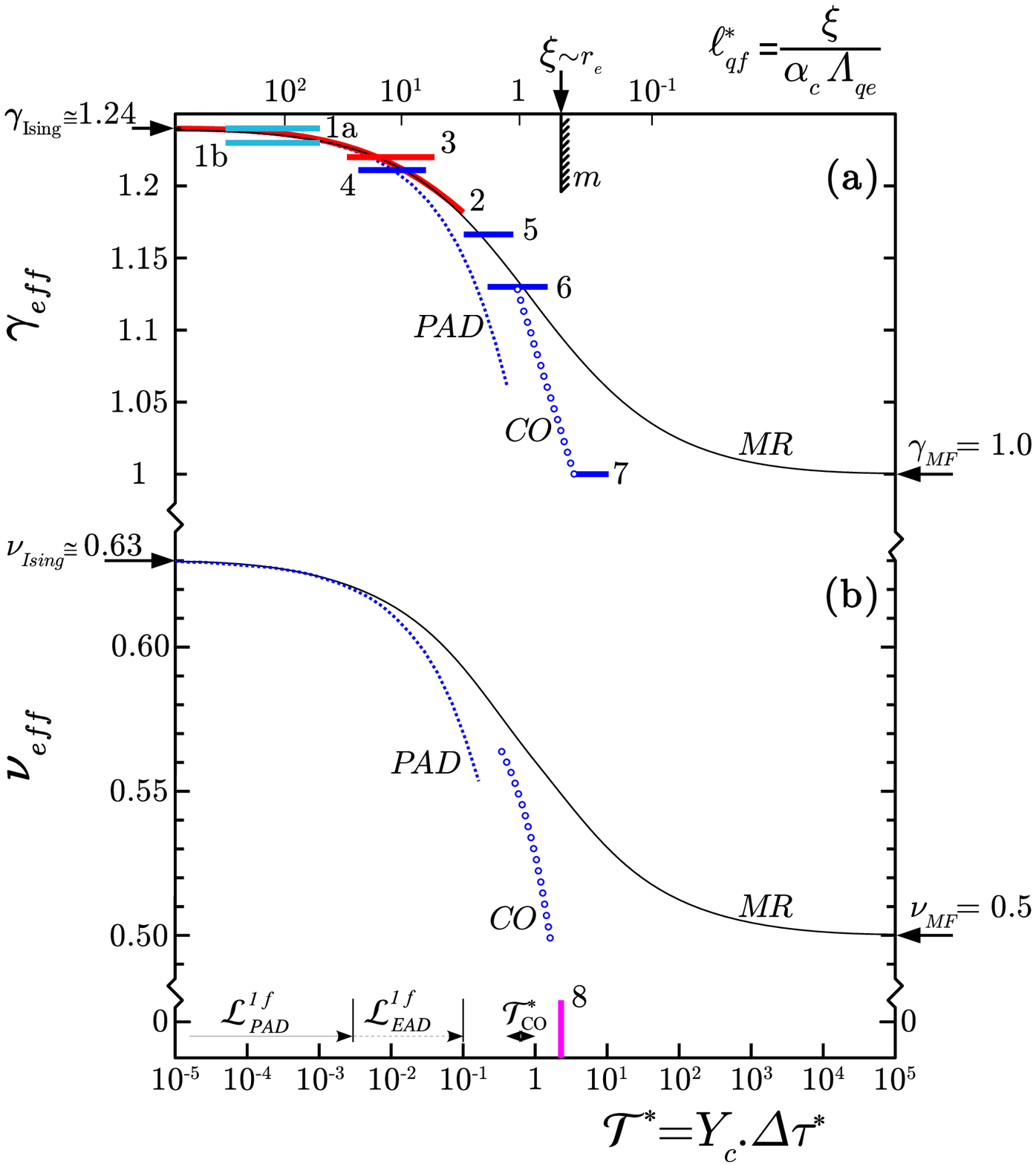}

\caption{(Color on line) Variations of the effective exponents $\gamma_{eff}$
(part a) and $\nu_{eff}$ (part b) as a function of the renormalized
thermal feld $\mathcal{T}^{*}$ (lower axis) or as a function of the
renormalized correlation length $\ell_{qf}^{\ast}$ (upper axis).
The full curves labelled $MR$ represent the $\gamma_{eff}$ and $\nu_{eff}$
theoretical variations obtained from the respective mean crossover
functions (see Ref. \cite{Garrabos 2005b} and text). The dashed curves
labelled $PAD$ correspond to the exponent variation in the preasymptotic
domain obtained from a Wegner expansion restricted to the first confluent
correction {[}see equation (\ref{two-term master equation}) and text
{]}, whose thermal field validity extends up to $\mathcal{L}_{PAD}^{1f}\lesssim3\,10^{-3}$
(see the corresponding arrow in $\mathcal{T}^{*}$ axis). In the extended
asymptotic domain (labelled $EAD$), the experimental master behavior
is well-represented by the theoretical crossover function {[}see equation
(\ref{lqf fitting equation}){]}, whose thermal field validity extends
up to $\mathcal{L}_{EAD}^{1f}\lesssim0.1$ (see the corresponding
arrow in $\mathcal{T}^{*}$ axis). The schematic curves labelled $CO$
and the curve $8$ (part b) correspond to the hypothetized \emph{effective}
crossover for pure fluids in a thermal field range $0.5\lesssim\mathcal{T}_{CO}^{*}\lesssim1$,
(see the corresponding double arrow in $\mathcal{T}^{*}$ axis). The
$1$-to-$7$ curves for $\gamma_{eff}$-variations refer {[}see also
Ref. (\cite{Bagnuls 1984b}){]} to experimental estimations of the
isothermal compressibility for xenon: $1a$ and $1b$) from interferometric
measurements (see Refs. \cite{Estler 1975} and \cite{Sengers 1978},
respectively); $2$) from light scattering measurements (see Ref.
\cite{Guttinger 1981}); $3$) from turbidity and ligth scattering
measurements (see Ref. \cite{Garrabos 1982}); $4$-to-$7$) from
$pVT$ measurement analysis (see Refs. \cite{Levelt-Sengers 1976,Garrabos 1982,Garrabos 1985}).
For the selected $pVT$ measurements see Refs. \cite{Habgood 1954,Michels 1954,Beattie 1951}.
For curve $m$ see the Figure 1 legend and text.}
\end{figure}

As illustrated by the curve labelled $CO$ in Figure 2a, the rough
estimate of the effective crossover around $\mathcal{T}_{CO}^{*}\approx0.5-1$
corresponds to the following noticeable differences in $\gamma_{eff}$-values
obtained from fitting analyses \cite{Levelt-Sengers 1976,Sengers 1978,Estler 1975,Garrabos 1982}
of the isothermal compressibility data of xenon obtained from $pVT$
measurements \cite{Habgood 1954,Michels 1954,Beattie 1951}: for $\mathcal{T}^{*}<\mathcal{T}_{CO}^{*}$,
$\gamma_{eff}$-values are always greater than $\frac{\gamma_{Ising}+\gamma_{MF}}{2}\cong1.12$
and increase to $\gamma_{Ising}\cong1.24$ when $\mathcal{T}^{*}\rightarrow0$;
for $\mathcal{T}^{*}\geq\mathcal{T}_{CO}^{*}$, $\gamma_{eff}$-value
is slightly constant and close to unity (see also the Figure 1 in
reference \cite{Bagnuls 1984b}). This general trend is observed whatever
the selected pure fluid, as already noted in references \cite{Garrabos 1982,Garrabos 1985}.
In spite of the difficulty in determining the precise shape of the
$\gamma_{eff}$-variation in this crossover range where $\ell_{qf}^{*}\lesssim1$,
we may expect to not observe a collapse onto an unique crossover curve
for different pure fluids (see also our analysis \cite{Garrabos 2002}
of crossover behavior for the renormalized order parameter of several
pure fluids in the non-homogeneous domain). As a matter of fact, the
massive renormalization scheme is not appropriate when $\ell_{qf}^{*}\lesssim1$.
We have illustrated this situation by the hypothetic curves labelled
$CO$ and $8$ in Figure 2b where a possible decrease of $\nu_{eff}$
to zero occurs crossing $\mathcal{T}_{CO}^{*}$. Such a zero-value
{[}significatively different from the (mean-field) $\frac{1}{2}$-value{]}
should be the result of an expected value $\ell_{qf}^{*}=const\cong\frac{1}{2}$
(see the limiting curve $m$, analogous to the one of Figure 1) related
to a constant value of the direct correlation length (however, as
already noted, $\ell_{qf}^{*}\approx\frac{1}{2}$ is not here a {}``master''
value whatever the pure fluid). At large temperature distance to the
critical point (i.e. for $\mathcal{T}\gg\mathcal{T}_{CO}$), such
a limit means that the direct correlation between interacting particles
at equilibrium position ($r_{e}\cong\frac{\alpha_{c}}{2}$) inside
the (short-ranged) critical interaction cell, can mainly contribute
to the local density fluctuations. We recall that, along the critical
isochore in the homogeneous phase, $\frac{1}{Z_{c}}$, i.e. the number
of particle in the critical interaction cell, and $\alpha_{c}$, i.e.
the size of the critical interaction cell, are two quantities which
are sligthly dependent on the temperature range (since the critical
isochore is closely a straight line in the $p;T$ diagram). Such a
conjectured microscopic situation is realistic and similar to the
results obtained from molecular numerical simulation of a Lennard-Jones
like fluid, precisely in the temperature range $T^{*}=\frac{k_{B}T}{\varepsilon_{LJ}}>2$
(which corresponds to $\mathcal{T}^{*}>1.5$ for the monoatomic rare
gases such as argon, krypton and xenon). In that situation, only the
first peak of the static structure factor is observed to be significant
in a reduced density {[}$\rho^{*}=\frac{\rho\left(\sigma_{LJ}\right)^{d}}{m_{\bar{p}}}\approx0.3${]}
range including the reduced critical density $\rho_{c}^{*}\approx0.3$
($\varepsilon_{LJ}$ and $\sigma_{LJ}$ are the two characteristics
parameters of the Lennard-Jones like potential). However, at such
{}``low'' density and {}``high'' temperature ranges, the associated
coordination number ($\lesssim4$) is too small to infer validity
of a mean-field approximation of the attractive interaction.

\section{Conclusion}

We provide an asymptotic description for the correlation length singular
behavior of the one-component fluid subclass. The master critical
crossover behavior of this subclass can be observed up to $\mathcal{T}^{\ast}\approx0.1$
(or $\ell_{qf}^{\ast}\approx3)$. The one-component fluid subclass
then corresponds to the simplest situation in the $\Phi_{d=3}^{4}\left(n=1\right)$-model
where the starting point for $u_{4}>0$ (in usual renormalized trajectories
\cite{Bagnuls 2000}), is certainly very close to the ideal RG trajectory
between the Gaussian and the Wilson-Fisher fixed points \cite{Bagnuls 2000}).
Such a supplementary constraint, is not a necessity in the field theory
framework \cite{Bagnuls 2002}. In a complementary work, we will show
that this constraint certainly takes origin in the description of
the critical point vicinity by \emph{finite} (linearized) thermodynamics,
which provides complete understanding for the \emph{master} thermodynamic
properties normalized at the volume scale $\left(\alpha_{c}\right)^{d}$
of the critical interaction cell.


\begin{thebibliography}{10}
\bibitem{particle mass}We admit that the molecular mass, $m_{\bar{p}}$, of each constitutive
fluid particle is a known quantity, in order to infer $N$ value from
total mass ($M=Nm_{\bar{p}})$ measurement of the amount of fluid
matter filling the container of measured total volume $V$.
\end{thebibliography}
\end{document}